\documentstyle[aps,preprint]{revtex}
\begin{document}
\title{Topological current of point defects and its bifurcation}
\author{Yishi Duan, Libin Fu\thanks{%
Corresponding author \mbox{} \mbox{} E-mail: itp2@lzu.edu.cn} and Hong Zhang}
\address{{\it Institute of Theoretical Physics, Lanzhou University, Lanzhou 730000, \\
P. R. China}}
\date{\today }
\maketitle

\begin{abstract}
From the topological properties of a three dimensional vector order
parameter, the topological current of point defects is obtained. One shows
that the charge of point defects is determined by Hopf indices and Brouwer
degrees. The evolution of point defects is also studied. One concludes that
there exist crucial cases of branch processes in the evolution of point
defects when the Jacobian $D(\frac \phi x)=0$.

{\bf PACS number(s): }03.40.Kf, 11.27.$+$d, 47.32.Cc, 02.40.Pc
\end{abstract}

\section{Introduction}

In recent years, an outstanding development in the theory of condensed
matter is to study the defect by topology. It has provided new insights and
spectacular predictions. In particular there has been progress on the study
the defects associated with a nonconserved $n$-component vector order
parameter field $\vec \phi (\vec r,t)$ \cite{Davis,BrayI}. In studying such
objects in physics arise to how one can define quantity like the density of
defects. Here comes the important conjecture proposed by Halperin, Liu and
Mazenko \cite{Halperin,Liu}, the density of such system has been wrote as 
\begin{equation}
\label{Liu}\rho =\delta (\vec \phi )D(\phi /x), 
\end{equation}
which is the fundamental equation for such problem.

In this paper, we will present a new topological current involved in
4-dimensional system by use of the $\phi $-mapping topological current
theory, which is important in studying the topological invariant and
structure of physics systems and has been used to study topological current
of magnetic monopole \cite{DuanGe}, topological string theory \cite{DuanLiu}%
, topological characteristics of dislocations and disclinations continuum 
\cite{DuanZhang}, topological structure of the defects of space-time in the
early universe as well as its topological bifurcation \cite
{DuanZhangFeng,DuanYangJiang}, topological structure of Gauss-Bonnet-Chern
theorem \cite{DuanMeng2,DuanLiYang} and topological structure of the London
equation in superconductor \cite{DuanZhangLi}. It is shown that the
topological charges of defects can be classified by Brower degree and Hopf
index of the $\phi $-mapping. It must be pointed out that the existence of
this topological current is inevitable and necessary, which carries all the
important topological properties of this physical system, and includes the
defect density even other conjectures given by Liu and Mazenko \cite{Liu}.
The further research shows that there exist the crucial cases of branch
process in the evolution of the point defects when the Jacobian $D(\frac \phi
x)=0$. We calculate the different branches of the worldlines of defects by
using the implicit function theorem. It is found that the worldlines of
point defects will split or merge at the critical points.

\section{The topological current of point defects}

Consider a 4-dimensional system with a nonconserved $3$-component vector
order parameter field $\vec \phi (\vec r,t)$. In order to compose a
topological current, we introduce a unit vector field 
\begin{equation}
\label{dwc}n^a=\frac{\phi ^a}{||\phi ||},\quad \ a=1,2,3 
\end{equation}
where 
$$
||\phi ||^2=\phi ^a\phi ^a. 
$$
$\vec \phi (x)$ is just the vector order parameter. The topological current
of this system is given by 
\begin{equation}
\label{currt}j^\mu =\frac 1{8\pi }\in ^{\mu \nu _1\upsilon _2\upsilon _3}\in
_{a_1a_2a_3}\partial _{\nu _1}n^{a_1}\partial _{\nu _2}n^{a_2}\partial _{\nu
_3}n^{a_3}. 
\end{equation}
It is clear that the topological current is identically conserved, i.e. 
\begin{equation}
\label{con}\partial _\mu j^\mu =0. 
\end{equation}
The topological charge density correspondingly will be defined as 
\begin{equation}
\label{den}\rho =j^0. 
\end{equation}
So, Eq. (\ref{con}) is just the continuity equation of this system.
Considering the following formula

$$
\partial _\mu n^a=\frac 1\phi \partial _\mu \phi ^a+\phi ^a\partial _\mu 
\frac 1\phi , 
$$
we can write Eq. (\ref{currt}) as

$$
j^\mu =\frac 1{8\pi }\in ^{\mu \mu _1\mu _2\mu _3}\in _{a_1a_2a_3}\partial
_{\mu _1}\phi ^a\partial _{\mu _2}\phi ^{a_2}\partial _{\mu _3}\phi ^{a_3}%
\frac \partial {\partial \phi ^a}\frac \partial {\partial \phi ^{a_1}}(\frac 
1{||\phi ||}). 
$$
If we define 
$$
\in ^{a_1a_2a_3}D^\mu (\frac \phi x)=\in ^{\mu \mu _1\mu _2\mu _3}\partial
_{\mu _1}\phi ^{a_1}\partial _{\mu _2}\phi ^{a_2}\partial _{\mu _3}\phi
^{a_3}, 
$$
in which $J^0(\frac \phi x)$ is just the usual 3-dimensional Jacobian
determinant 
$$
D^0(\frac \phi x)=D(\frac \phi x)=\left( 
\begin{array}{ccc}
\frac{\partial \phi ^1}{\partial x^1} & \frac{\partial \phi ^1}{\partial x^2}
& \frac{\partial \phi ^1}{\partial x^3} \\ \frac{\partial \phi ^2}{\partial
x^1} & \frac{\partial \phi ^2}{\partial x^2} & \frac{\partial \phi ^2}{%
\partial x^3} \\ \frac{\partial \phi ^3}{\partial x^1} & \frac{\partial \phi
^3}{\partial x^2} & \frac{\partial \phi ^3}{\partial x^3} 
\end{array}
\right) , 
$$
and make use of the 3-dimensional Laplacian Green's function relation 
$$
\bigtriangleup _\phi (\frac 1{||\phi ||})=-4\pi \delta ^3(\vec \phi ) 
$$
where $\bigtriangleup _\phi $ is the $3$-dimensional Laplacian operator in $%
\phi $ space, we do obtain the $\delta $ function like current 
\begin{equation}
\label{5}j^\mu =\delta ^3(\vec \phi )D^\mu (\frac \phi x). 
\end{equation}
When $\mu =0$, we know that the density of $j^\mu $ can be written as

\begin{equation}
\label{7}\rho =\delta ^3(\vec \phi )D(\frac \phi x). 
\end{equation}

From this expression, we find that $J^\mu $ does not vanish only when $\vec 
\phi =0$, i.e. 
\begin{equation}
\label{eqution}\phi
^a(t,x^1,x^2,x^3)=0,\,\,\,\,\,\,\,\,\,\,\,\,\,\,\,\,\,a=1,2,3. 
\end{equation}
Suppose that the vector field $\vec \phi $ possesses $l$ zeroes (i.e., there
exist $l$ point defects) denoted as $z_i$($i=1,...,l$ ). According to the
implicit function theorem \cite{Goursat}, when the zeros points $\vec z_i$
are the regular points of $\vec \phi $ which requires the Jacobian 
\begin{equation}
\label{non}D(\frac \phi x)|_{z_i}=D^0(\frac \phi x)|_{z_i}\neq 0. 
\end{equation}
the solutions of Eq. (\ref{eqution}) can be generally obtained: 
\begin{equation}
\label{solu}\vec x=\vec z_i(t)\,\,\,\,\,\,\,\,\,\,\,\,\,\,\,\,\,\,%
\,i=1,2,...,l\,\,\,\,\,\,\,\,\,\,\,\, 
\end{equation}
which represent the worldlines of $l$ point defects moving in space. From
Eq. (\ref{eqution}), it is easy to prove that 
\begin{equation}
\label{velo}D^\mu (\frac \phi x)|_{z_i}=D(\frac \phi x)|_{z_i}\frac{dx^\mu }{%
dt}. 
\end{equation}
So, the topological current (\ref{currt}) can be rigorously expressed as 
\begin{equation}
\label{p1}J^\mu =\delta ^3(\vec \phi )D(\frac \phi x)\frac{dx^\mu }{dt} 
\end{equation}
As we proved in Ref. \cite{DuanYangJiang}, we have 
\begin{equation}
\label{curr}J^\mu =\sum_{i=1}^l\beta _i\eta _i\delta ^3(\vec x-\vec z_i)%
\frac{dx^\mu }{dt}|_{z_i}, 
\end{equation}
where the positive integer $\beta _i$ is the Hopf index and $\eta
_i=signD(\phi /x)_{z_i}=\pm 1$ is Brouwer degree\cite{DuanGe,DuanZhang}.
According to Eq. (\ref{curr}), the density of topological charge can be
rewritten as 
\begin{equation}
\label{densit}\rho =J^0=\sum_{i=1}^l\beta _i\eta _i\delta ^3(\vec x-\vec z%
_i). 
\end{equation}
It is clearly that the inner structure of this system is characterized by
Hopf indices $\beta _i$ and Brouwer degrees $\eta _i$, which are topological
invariants. From discussions above, we see that the density $\rho (x)$ is
similar to a system of $l$ classical point-like particles with topological
charge $\beta _i\eta _i$ moving in the 4-dimensional space-time. The
topological charge $\beta _i\eta _i$ is also called the topological charge
of the {\it i}th point defect and $\rho (x)$ can be regarded as the density
of defects. And , solution (\ref{solu}) can be regarded as the trajectory of
the {\it i}th defect. From formula (\ref{densit}), we obtain that the total
charge of the system 
\begin{equation}
\label{end}Q=\int \rho (x)d^3x=\sum_{i=1}^l\beta _i\eta _i. 
\end{equation}

The result (\ref{7}) also has been carried out by Halperin \cite{Halperin},
and exploited by Liu and Mazenko \cite{Liu}: the first ingredient is the
rather obvious result 
$$
\sum\limits_\alpha \delta (\vec{r}-\vec{r}_\alpha (t))=\delta (\vec{\phi}(%
\vec{r},t))\mid D(\frac \phi x)\mid 
$$
where the second factor on the right-hand side is just the Jacobian of the
transformation from the variable $\vec{\phi}$ to $\vec{r}$. This is combined
with the less obvious result

$$
\eta _\alpha =sgnD(\phi /x)\mid _{\vec r_\alpha } 
$$
to give 
\begin{equation}
\label{Liu2}\rho (\vec r,t)=\sum\limits_\alpha \eta _\alpha \delta (\vec r-%
\vec r_\alpha (t))=\delta (\vec \phi )D(\phi /x). 
\end{equation}
Here we see that the result (\ref{Liu2}) obtained by Halperin, Liu and
Mazenko is not complete. They did not considering the cases $\beta _l\neq 1$
and $D(\phi /x)=0$, i.e., $\eta _l$ is indefinite. It is interesting to
discuss what will happen and what does it correspond to in physics when $%
D(\phi /x)=0$.

\section{The bifurcation of worldlines of point defects}

As being discussed before, the zeros of the smooth vector $\vec \phi $
(locations of defects) play important roles in studying the evolution of
point defect. In this section, we will study the properties of the zero
points, in other words, the properties of the solutions of following
equations 
\begin{equation}
\label{q1}\left\{ 
\begin{array}{c}
\phi ^1(x^0,x^1,x^2,x^3)=0 \\ 
\phi ^2(x^0,x^1,x^2,x^3)=0 \\ 
\phi ^3(x^0,x^1,x^2,x^3)=0 
\end{array}
\right. . 
\end{equation}
As we knew before, if the Jacobian determinant 
$$
D(\frac \phi x)|_{z_i}=D^0(\frac \phi x)|_{z_i}\neq 0 
$$
we will have the isolated solutions (\ref{solu}) of Eq. (\ref{q1}). The
isolated solutions are called regular points. It is easy to see that the
results in section II is based on this condition. However, when this
condition fails, the above results will change in some way, and will lead to
the branch process of topological density and give rise to the bifurcation.
We denote one of zero points as $(t^{*},\vec z_i)$.

\subsection{the branch process at a limit point}

It is well-known that when the Jacobian $D(\frac \phi x)|_{(t^{*},\vec z%
_i)}=0$, the usual implicit function theorem is no use. But if the Jacobian 
$$
D^1(\frac \phi x)|_{(t^{*},\vec z_i)}=\frac{\partial (\phi ^1,\phi ^2,\phi
^3)}{\partial (t,x^2,x^3)}|_{(t^{*},\vec z_i)}\neq 0, 
$$
we can use the Jacobian $D^1(\frac \phi x)|_{(t^{*},\vec z_i)}$ instead of $%
D(\frac \phi x)|_{(t^{*},\vec z_i)}$, for the purpose of using the implicit
function theorem \cite{Goursat}. Then we have an unique solution of Eqs. (%
\ref{q1}) in the neighborhood of the point ($t^{*},\vec z_i$) 
$$
t=t(x^1) 
$$
\begin{equation}
\label{q4}x^i=x^i(x^1)\qquad i=2,3. 
\end{equation}
with $t^{*}=t(x^1)$. And we call the critical points ($t^{*},\vec z_i$) the
limit points. In the present case, it is easy to know that 
\begin{equation}
\label{wqd}\frac{dx^1}{dt}|_{(t^{*},\vec z_i)}=\frac{D^1(\frac \phi x%
)|_{(t^{*},\vec z_i)}}{D(\frac \phi x)|_{(t^{*},\vec z_i)}}=\infty 
\end{equation}
i.e. 
$$
\frac{dt}{dx^1}|_{(t^{*},\vec z_i)}=0. 
$$
Then we have the Taylor expansion of Eq. (\ref{q4}) at the point $\left(
t^{*},\vec z_i\right) $%
$$
t=t^{*}+\frac{dt}{dx^1}|_{(t^{*},\vec z_i)}(x^1-z_i^1)+\frac 12\frac{d^2t}{%
(dx^1)^2}|_{(t^{*},\vec z_i)}(x^1-z_i^1)^2 
$$
$$
=t^{*}+\frac 12\frac{d^2t}{(dx^1)^2}|_{(t^{*},\vec z_i)}(x^1-z_i^1)^2. 
$$
Therefore 
\begin{equation}
\label{q5}t-t^{*}=\frac 12\frac{d^2t}{(dx^1)^2}|_{(t^{*},\vec z%
_i)}(x^1-z_i^1)^2 
\end{equation}
which is a parabola in the $x^1-t$ plane. From Eq. (\ref{q5}), we can obtain
two worldlines of two point defects $x_1^1(t)$ and $x_2^1(t)$, which give
the branch solutions of the system (\ref{q1}). If $\frac{d^2t}{(dx^1)^2}%
|_{(t^{*},\vec z_i)}>0$, we have the branch solutions for $t>t^{*}$,
otherwise, we have the branch solutions for $t>t^{*}$. It is clear that
these two cases are related to the origin and annihilation of point defects.
Since the topological charge of point defect is identically conserved, the
topological charge of these two must be opposite at the zero point, i.e. 
\begin{equation}
\label{chargeI}\beta _{i1}\eta _{i1}=-\beta _{i2}\eta _{i2}. 
\end{equation}

One of the result of Eq. (\ref{wqd}), that the velocity of point defects is
infinite when they are annihilating, agrees with that obtained by Bray \cite
{Bray} who has a scaling argument associated with point defects final
annihilation which leads to large velocity tail. From Eq. (\ref{wqd}), we
also get a new result that the velocity of point defects is infinite when
they are generating, which is gained only from the topology of the three
dimensional vector order parameter.

For a limit point, it also requires the $D^1(\frac \phi x)|_{(t^{*},\vec z%
_i)}\neq 0.$\ As to a bifurcation point \cite{Kubicek}, it must satisfy a
more complement condition. This case will be discussed in the following
subsections in detail.

\subsection{the branch process at a bifurcation point}

In this subsection, we have the restrictions of the system (\ref{q1}) at the
bifurcation point $(t^{*},\vec z_i)$: 
\begin{equation}
\label{bifa12}\left\{ 
\begin{array}{c}
D( 
\frac \phi x)|_{(t^{*},\vec z_i)}=0 \\ D^1(\frac \phi x)|_{(t^{*},\vec z%
_i)}=0 
\end{array}
\right. 
\end{equation}

These will lead to an important fact that the function relationship between $%
t$ and $x^1$ is not unique in the neighborhood of the bifurcation point $(%
\vec z_i,t^{*})$. It is easy to see from equation 
\begin{equation}
\label{bifa13}\frac{dx^1}{dt}\mid _{(t^{*},\vec z_i)}=\frac{D^1(\frac \phi x%
)|_{(t^{*},\vec z_i)}}{D(\frac \phi x)|_{(t^{*},\vec z_i)}} 
\end{equation}
which under the restraint (\ref{bifa12}) directly shows that the direction
of the worldlines of point defects is indefinite at the point $(\vec z%
_i,t^{*})$. This is why the very point $(\vec z_i,t^{*})$ is called a
bifurcation point of the system (\ref{q1}).

Next, we will find a simple way to search for the different directions of
all branch curves at the bifurcation point. Assume that the bifurcation
point $(\vec z_i,t^{*})$ has been found from Eqs. (\ref{q1}) and (\ref
{bifa12}). We know that, at the bifurcation point $(\vec z_i,t^{*})$, the
rank of the Jacobian matrix $[\frac{\partial \phi }{\partial x}]$ is smaller
than $3$. First, we suppose the rank of the Jacobian matrix $[\frac{\partial
\phi }{\partial x}]$ is $2$ (the case of a more smaller rank will be
discussed later). Suppose that the $2\times 2$ submatrix $J_1(\frac \phi x%
)\, $ is 
\begin{equation}
\label{bifa14}J_1(\frac \phi x)=\left( 
\begin{array}{cc}
\frac{\partial \phi ^1}{\partial x^2} & \frac{\partial \phi ^1}{\partial x^3}
\\ \frac{\partial \phi ^2}{\partial x^2} & \frac{\partial \phi ^2}{\partial
x^3} 
\end{array}
\right) , 
\end{equation}
and its determinant $D^1(\frac \phi x)$ does not vanish. The implicit
function theorem says that there exist one and only one function relation 
\begin{equation}
\label{bifa15}x^i=f^i(x^1,t),\quad i=2,3 
\end{equation}
We denoted the partial derivatives as 
$$
f_1^i=\frac{\partial f^i}{\partial x^1};\,\quad \,f_t^i=\frac{\partial f^i}{%
\partial t};\quad f_{11}^i=\frac{\partial ^2f^i}{\partial x^1\partial x^1}%
;\quad f_{1t}^i=\frac{\partial ^2f^i}{\partial x^1\partial x^t};\quad
f_{tt}^i=\frac{\partial f^i}{\partial x^t\partial x^t} 
$$
From Eqs. (\ref{q1}) and (\ref{bifa15}) we have for $a=1,2,3$ 
\begin{equation}
\label{bifa16}\phi ^a=\phi ^a(x^1,f^2(x^1,t),f^3(x^1,t),t)=0 
\end{equation}
which give 
\begin{equation}
\label{bifa18}\frac{\partial \phi ^a}{\partial x^1}=\phi _1^a+\sum_{j=2}^3%
\frac{\partial \phi ^a}{\partial f^j}\frac{\partial f^j}{\partial x^1}=0 
\end{equation}
\begin{equation}
\label{bifa19}\frac{\partial \phi ^a}{\partial t}=\phi _t^a+\sum_{j=2}^3%
\frac{\partial \phi ^a}{\partial f^j}\frac{\partial f^j}{\partial t}=0. 
\end{equation}
from which we can get the first order derivatives of $f^i$: $\,f_1^i$ and $%
f_t^i$. Differentiating Eq. (\ref{bifa18}) with respect to $x^1$ and $t$
respectively we get 
\begin{equation}
\label{bifa21}\sum_{j=2}^3\phi _j^af_{11}^j=-\sum_{j=2}^3[2\phi
_{j1}^af_1^j+\sum_{k=2}^3(\phi _{jk}^af_1^k)f_1^j]-\phi _{11}^a\quad \quad
a=1,2,3 
\end{equation}
\begin{equation}
\label{bifa23}\sum_{j=2}^3\phi _j^af_{1t}^j=-\sum_{j=2}^3[\phi
_{jt}^af_1^j+\phi _{j1}^af_t^j+\sum_{k=2}^3(\phi _{jk}^af_t^k)f_1^j]-\phi
_{1t}^a\quad \quad a=1,2,3 
\end{equation}
And the differentiation of Eq. (\ref{bifa19}) with respect to $t$ gives 
\begin{equation}
\label{bifa24}\sum_{j=2}^3\phi _j^af_{tt}^j=-\sum_{j=2}^3[2\phi
_{jt}^af_t^j+\sum_{k=2}^3(\phi _{jk}^af_t^k)f_t^j]-\phi _{tt}^a\quad \quad
\quad a=1,2,3 
\end{equation}
where 
\begin{equation}
\label{bifa25}\phi _{jk}^a=\frac{\partial ^2\phi ^a}{\partial x^j\partial x^k%
},\quad \quad \phi _{jt}^a=\frac{\partial ^2\phi ^a}{\partial x^j\partial t}%
. 
\end{equation}
The differentiation of Eq. (\ref{bifa19}) with respect to $x^1$ gives the
same expression as Eq. (\ref{bifa23}). By making use of the Gaussian
elimination method to Eqs. (\ref{bifa23}), (\ref{bifa23}) and (\ref{bifa24})
we can find the second order derivatives $f_{11}^i$, $f_{1t}^i$ and $%
f_{tt}^i $. The above discussion does no matter to the last component $\phi
^3(\vec x,t)$. In order to find the different values of $dx^1/dt$ at the
bifurcation point $(\vec z_i,t^{*})$, let us investigate the Taylor
expansion of $\phi ^3(\vec x,t)$ in the neighborhood of $(\vec z_i,t^{*})$.
Substituting Eq. (\ref{bifa15}) into $\phi ^3(\vec x,t)$ we have the
function of two variables $x^1$ and $t$ 
\begin{equation}
\label{bifa26}F(x^1,t)=\phi ^3(x^1,f^2(x^1,t),f^3(x^1,t),t) 
\end{equation}
which according to Eq. (\ref{q1}) must vanish at the bifurcation point 
\begin{equation}
\label{bifa27}F(z_i^1,t^{*})=0. 
\end{equation}
From Eq. (\ref{bifa26}) we have the first order partial derivatives of $%
F(x^1,t)$ 
\begin{equation}
\label{bifa28}\frac{\partial F}{\partial x^1}=\phi _1^3+\sum_{j=2}^3\phi
_j^3f_1^j,\quad \quad \frac{\partial F}{\partial t}=\phi
_t^3+\sum_{j=2}^3\phi _j^3f_t^j. 
\end{equation}
Using Eqs. (\ref{bifa18}) and (\ref{bifa19}) the first equation of (\ref
{bifa12}) is expressed as 
\begin{equation}
\label{bifa29}D(\frac \phi x)|_{(\vec z_i,t^{*})}=\left| 
\begin{array}{ccc}
-\sum\limits_{j=2}^3\phi _j^1f_1^j & \phi _2^1 & \phi _3^1 \\ 
-\sum\limits_{j=2}^3\phi _j^2f_1^j & \phi _2^2 & \phi _3^2 \\ 
\phi _1^3 & \phi _2^3 & \phi _3^3 
\end{array}
\right| _{(\vec z_i,t^{*})}=0 
\end{equation}
which by Cramer's rule can be written as 
$$
D(\frac \phi x)|_{(\vec z_i,t^{*})}=\frac{\partial F}{\partial x^1}\det J_1(%
\frac \phi x)|_{((\vec z_i,t^{*})}=0 
$$
Since $\det J_1(\frac \phi x)|_{(\vec z_i,t^{*})}\neq 0$, the above equation
gives 
\begin{equation}
\label{bifa31}\frac{\partial F}{\partial x^1}\mid _{(\vec z_i,t^{*})}=0. 
\end{equation}
With the same reasons, we have 
\begin{equation}
\label{bifa32}\frac{\partial F}{\partial t}\mid _{(\vec z_i,t^{*})}=0. 
\end{equation}
The second order partial derivatives of the function $F$ are easily to find
out to be 
\begin{equation}
\label{bifa33}\frac{\partial ^2F}{(\partial x^1)^2}=\phi
_{11}^3+\sum\limits_{j=2}^3[2\phi _{1j}^3f_1^j+\phi
_j^3f_{11}^j+\sum\limits_{k=2}^3(\phi _{kj}^3f_1^k)f_1^j] 
\end{equation}
\begin{equation}
\label{bifa34}\frac{\partial ^2F}{\partial x^1\partial t}=\phi
_{1t}^3+\sum\limits_{j=2}^3[\phi _{1j}^3f_t^j+\phi _{tj}^3f_1^j+\phi
_j^3f_{1t}^j+\sum\limits_{k=2}^3(\phi _{jk}^3f_t^k)f_1^j] 
\end{equation}
\begin{equation}
\label{bifa35}\frac{\partial ^2F}{\partial t^2}=\phi
_{tt}^3+\sum\limits_{j=2}^3[2\phi _{jt}^3f_t^j+\phi
_j^3f_{tt}^j+\sum\limits_{k=2}^3(\phi _{jk}^3f_t^k)f_t^j] 
\end{equation}
which at $(\vec z_i,t^{*})$ are denoted by 
\begin{equation}
\label{bifa36}A=\frac{\partial ^2F}{(\partial x^1)^2}\mid _{(\vec z%
_i,t^{*})},\quad \quad B=\frac{\partial ^2F}{\partial x^1\partial t}\mid _{(%
\vec z_i,t^{*})},\ \quad \quad C=\frac{\partial ^2F}{\partial t^2}\mid _{(%
\vec z_i,t^{*})}. 
\end{equation}
Then taking notice of Eqs. ( \ref{bifa27}), (\ref{bifa31}), (\ref{bifa32})
and (\ref{bifa36}) the Taylor expansion of $F(x^1,t)$ in the neighborhood of
the bifurcation point $(\vec z_i,t^{*})$ can be expressed as 
\begin{equation}
\label{bifa37}F(x^1,t)=\frac 12A(x^1-z_i^1)^2+B(x^1-z_i^1)(t-t^{*})+\frac 12%
C(t-t^{*})^2 
\end{equation}
which by Eq. (\ref{bifa26}) is the expression of $\phi ^3(\vec x,t)$ in the
neighborhood of $(\vec z_i,t^{*}).$ The expression (\ref{bifa37}) is
reasonable, which shows that at the bifurcation point $(\vec z_i,t^{*})$ one
of Eqs. (\ref{q1}), $\phi ^3(\vec x,t)=0,$ is satisfied, i.e. 
\begin{equation}
\label{bifa38}A(x^1-z_i^1)^2+2B(x^1-z_i^1)(t-t^{*})+C(t-t^{*})^2=0. 
\end{equation}
Dividing Eq. (\ref{bifa38}) by $(t-t^{*})^2$ and taking the limit $%
t\rightarrow t^{*}$ as well as $x^1\rightarrow z_i^1$ respectively we get 
\begin{equation}
\label{bifb38}A(\frac{dx^1}{dt})^2+2B\frac{dx^1}{dt}+C=0. 
\end{equation}
In the same way we have 
\begin{equation}
\label{bifa39}C(\frac{dt}{dx^1})^2+2B\frac{dt}{dx^1}+A=0. 
\end{equation}
where $A$, $B$ and $C$\ are three constants. The solutions of Eq. (\ref
{bifb38}) or Eq. (\ref{bifa39}) give different directions of the branch
curves (worldlines of point defects) at the bifurcation point. There are
four possible cases, which will show the physical meanings of the
bifurcation points.

Case 1 ($A\neq 0$): For $\Delta =4(B^2-AC)>0$\ from Eq. (\ref{bifb38}) we
get two different directions of the velocity field of point defects 
\begin{equation}
\label{case1}\frac{dx^1}{dt}\mid _{1,2}=\frac{-B\pm \sqrt{B^2-AC}}A, 
\end{equation}
\ where two worldlines of two point defects intersect with different
directions at the bifurcation point. This shows that two point defects
encounter and then depart at the bifurcation point.

Case 2 ($A\neq 0$): For $\Delta =4(B^2-AC)=0$\ from Eq. (\ref{bifb38}) we
get only one direction of the velocity field of point defects 
\begin{equation}
\label{case2}\frac{dx^1}{dt}\mid _{1,2}=-\frac BA. 
\end{equation}
\ which includes three important cases. (a) Two worldlines tangentially
contact, i.e. two point defects tangentially encounter at the bifurcation
point. (b) Two worldlines merge into one worldline, i.e. two point defects
merge into one point defect at the bifurcation point. (c) One worldline
resolves into two worldlines, i.e. one point defect splits into two point
defects at the bifurcation point.

Case 3 ($A=0,\,C\neq 0$): For $\Delta =4(B^2-AC)=0$\ from Eq. (\ref{bifa39})
we have 
\begin{equation}
\label{case3}\frac{dt}{dx^1}\mid _{1,2}=\frac{-B\pm \sqrt{B^2-AC}}C=0,\quad -%
\frac{2B}C. 
\end{equation}
\ There are two important cases: (a) One worldline resolves into three
worldlines, i.e. one point defect splits into three point defects at the
bifurcation point. (b) Three worldlines merge into one worldline, i.e. three
point defects merge into one point defect at the bifurcation point.

Case 4 ($A=C=0$): Eq. (\ref{bifb38}) and Eq. (\ref{bifa39}) gives
respectively 
\begin{equation}
\label{case4}\frac{dx^1}{dt}=0,\quad \quad \frac{dt}{dx^1}=0. 
\end{equation}
\ This case is obvious similar to case 3.

The above solutions reveal the evolution of the point defects. Besides the
encountering of the point defects, i.e. two point defects encounter and then
depart at the bifurcation point along different branch curves, it also
includes spliting and merging of point defects. When a multicharged point
defect moves through the bifurcation point, it may split into several point
defects along different branch curves. On the contrary, several point
defects can merge into one point defect at the bifurcation point. The
identical conversation of the topological charge shows the sum of the
topological charge of final point defect(s) must be equal to that of the
initial point defect(s) at the bifurcation point, i.e., 
\begin{equation}
\label{chargeII}\sum_f\beta _{l_f}\eta _{l_f}=\sum_i\beta _{l_i}\eta _{l_i} 
\end{equation}
for fixed $l$. Furthermore, from above studies, we see that the generation,
annihilation and bifurcation of point defects are not gradual changes, but
start at a critical value of arguments, i.e. a sudden change.

\subsection{the branch process at a higher degenerated point}

In above subsection, we have studied the case that the rank of the Jacobian
matrix $[\frac{\partial \phi }{\partial x}]$ of Eqs. (\ref{q1}) is $2=3-1$.
In this subsection, we consider the case that the rank of the Jacobian
matrix is $1=3-2$. Let the $J_2(\frac \phi x)=\frac{\partial \phi ^1}{%
\partial x^1}\,$ and suppose that $\det J_2\neq 0$. With the same reasons of
obtaining Eq. (\ref{bifa15}), we can have the function relations 
\begin{equation}
\label{bifa45}x^3=f^3(x^1,x^2,t)\quad . 
\end{equation}
Substituting relations (\ref{bifa45}) into the last two equations of (\ref
{q1}), we have following two equations with three arguments $x^1,x^2,t$ 
\begin{equation}
\label{bifa46}\left\{ 
\begin{array}{l}
F_1(x^1,x^2,t)=\phi ^2(x^1,x^2,f^3(x^1,x^2,t),t)=0 \\ 
F_2(x^1,x^2,t)=\phi ^3(x^1,x^2,f^3(x^1,x^2,t),t)=0. 
\end{array}
\right. 
\end{equation}
Calculating the partial derivatives of the function $F_1$ and $F_2$ with
respect to $x^1$, $x^2$ and $t$, taking notice of Eq. (\ref{bifa45}) and
using six similar expressions to Eqs. (\ref{bifa31}) and (\ref{bifa32}),
i.e. 
\begin{equation}
\label{bifa48}\frac{\partial F_j}{\partial x^1}\mid _{(\vec{z}_i,t^{*})}=0,\
\quad \quad \frac{\partial F_j}{\partial x^2}\mid _{(\vec{z}_i,t^{*})}=0,\
\quad \quad \frac{\partial F_j}{\partial t}\mid _{(\vec{z}_i,t^{*})}=0,\
\quad \quad j=1,2, 
\end{equation}
we have the following forms of Taylor expressions of $F_1$ and $F_2$ in the
neighborhood of $(\vec{z}_i,t^{*})$ 
$$
F_j(x^1,x^2,t)\approx
A_{j1}(x^1-z_i^1)^2+A_{j2}(x^1-z_i^1)(x^2-z_i^2)+A_{j3}(x^1-z_i^1) 
$$
$$
\times
(t-t^{*})+A_{j4}(x^2-z_i^2)^2+A_{j5}(x^2-z_i^2)(t-t^{*})+A_{j6}(t-t^{*})^2=0 
$$
\begin{equation}
\label{bifa49}j=1,2. 
\end{equation}
In case of $A_{j1}\neq 0$ and $A_{j4}\neq 0$, dividing Eq. (\ref{bifa49}) by 
$(t-t^{*})^2$ and taking the limit $t\rightarrow t^{*}$, we obtain two
quadratic equations of $\frac{dx^1}{dt}$ and $\frac{dx^2}{dt}$ 
\begin{equation}
\label{bifa50}A_{j1}(\frac{dx^1}{dt})^2+A_{j2}\frac{dx^1}{dt}\frac{dx^2}{dt}%
+A_{j3}\frac{dx^1}{dt}+A_{j4}(\frac{dx^2}{dt})^2+A_{j5}\frac{dx^2}{dt}%
+A_{j6}=0 
\end{equation}
$$
j=1,2. 
$$
Eliminating the variable $dx^1/dt$, we obtain a equation of $dx^2/dt$ in the
form of a determinant 
\begin{equation}
\label{bifa51}\left| 
\begin{array}{cccc}
A_{11} & A_{12}v+A_{23} & A_{14}v^2+A_{15}v+A_{16} & 0 \\ 
0 & A_{11} & A_{12}v+A_{13} & A_{14}v^2+A_{15}v+A_{16} \\ 
A_{21} & A_{22}v+A_{23} & A_{24}v^2+A_{25}v+A_{26} & 0 \\ 
0 & A_{21} & A_{22}v+A_{23} & A_{24}v^2+A_{25}v+A_{26} 
\end{array}
\right| =0 
\end{equation}
where $v=dx^2/dt$, that is a {\it 4}th order equation of $dx^2/dt$ 
\begin{equation}
\label{bifa52}a_0(\frac{dx^2}{dt})^4+a_1(\frac{dx^2}{dt})^3+a_2(\frac{dx^2}{%
dt})^2+a_3(\frac{dx^2}{dt})+a_4=0. 
\end{equation}
Therefore we get different directions of the worldlines of point defects at
the higher degenerated point bifurcation point. The number of different
branch curves is at most four.

At the end of this section, we conclude that in our theory of point defects
there exist the crucial case of branch process \cite{DuanLiYang}. Besides
the encountering of the point defects, i.e. two point defects encounter and
then depart at the bifurcation point along different worldlines, it also
includes spliting and merging of point defects. When a multicharged point
defect moves through the bifurcation point, it may split into several point
defects along different worldlines. On the contrary, several point defects
can merge into one point defect at the bifurcation point. Since the
topological charges of point defects is identically conserved (\ref{con}),
the sum of the topological charge of final point defect(s) must be equal to
that of the initial point defect(s) at the bifurcation point.

\section{Conclusions}

We have studied the evolution of the point defects of a three dimensional
vector order parameter by making use of the $\phi $ mapping topological
current theory. We conclude that there exist crucial cases of branch
processes in the evolution of the point defects. This means that the point
defects generate or annihilate at the limit points and encounter, split or
merge at the bifurcation points of the three dimensional vector order
parameter, which shows that the point defects system is unstable at these
branch points. There are two restrictions of the evolution of point defects
in this paper. One restriction is the conservation of the topological charge
of the point defects during the branch process, the other restriction is the
number of different directions of the worldlines of point defects is at most
four at the bifurcation points. At last, we would like to point out that all
the results in this paper are obtained from the viewpoint of topology
without using any particular models or hypothesis.

\section*{Acknowledgment}

This work was supported by the National Natural Science Foundation of the
People's Republic of China.

\end{document}